\pdfoutput=1 
\documentclass[%
 reprint,
 amsmath,amssymb,
 aps,
 prl
 longbibliography,
]{revtex4-2}

\usepackage[pdftex]{graphicx} 
\usepackage{dcolumn}
\usepackage{bm}
\usepackage[T1]{fontenc}
\usepackage[pdftex]{color, hyperref}
\hypersetup{
colorlinks = true,
linkcolor = blue,
citecolor = blue,
}

\usepackage{color}

\newcommand{\parhyphen}[1][1em]{
  {\normalfont\normalsize\bfseries\hspace*{-1em}}\hspace*{#1}---\hspace*{#1}\ignorespaces}

\begin{document}
\footnotetext[1]{Here, we do not consider the case that the nonlinear term itself dissipates energy, as in the Onsager ``ideal turbulence'' theory \cite{Onsager_1949,Eyink_Review}.}
\footnotetext[2]{See Supplemental Material at [URL will be inserted by publisher] for the case $d=3$.}
\footnotetext[3]{See Supplemental Material at [URL will be inserted by publisher] for the scale locality.}


\title{A Simple XY Model for Cascade Transfer}

\author{Tomohiro Tanogami}
\author{Shin-ichi Sasa}%
\affiliation{%
Department of Physics, Kyoto University, Kyoto 606-8502, Japan
}%




\date{\today}
\begin{abstract}
We propose a modified XY model in which cascade transfer emerges from spatially local interactions, where the spin corresponds to the ``velocity'' of a turbulent field. 
For this model, we theoretically calculate the scale-to-scale energy flux and the equal time correlation function in $d$ dimensions. 
The result indicates an inverse energy cascade with the non-Kolmogorov energy spectrum proportional to $k^{-3}$. 
We also numerically confirm the result for the cases $d=2$ and $d=3$. 
We thus conclude that the cascade transfer in our model represents a different universality class from standard fluid turbulence.
\end{abstract}

\pacs{Valid PACS appear here}

\maketitle
\paragraph*{Introduction.}\parhyphen[0pt]
Many phenomena in nature can be regarded as cooperative phenomena in the sense that they emerge from interactions between many components.
Although such interactions are complicated, the resulting cooperative phenomena themselves are often universal regardless of the details of the interaction, allowing for a phenomenological understanding.
Therefore, if we are only interested in the universal aspect of a certain phenomenon, it is sufficient to investigate the simplest model that can describe the phenomenon.
Simple models have provided us phenomenological perspectives with which to understand various phenomena such as critical phenomena \cite{goldenfeld2018lectures}, phase separation \cite{oono1987computationally,oono1988study}, directed percolation \cite{hinrichsen2000non}, surface growth \cite{barabasi1995fractal,takeuchi2018appetizer}, and flocking \cite{ramaswamy2010mechanics}.

One of the most extreme examples in which complicated interaction between many degrees of freedom plays a central role is cascade transfer \cite{Frisch}.
This is the phenomenon that an inviscid conserved quantity, such as energy or enstrophy, is transferred conservatively from large (small) to small (large) scales.
In the scale range where the cascade transfer occurs---the inertial range---the scaling of the distribution of the conserved quantity is governed by the corresponding conserved scale-to-scale flux.
In other words, cascade transfer underlies the remarkable universality of the scaling.
In fluid turbulence, for example, the energy spectrum follows the Kolmogorov spectrum $E(k)\propto k^{-5/3}$ independently of the details of the initial/boundary conditions or the mechanism of external stirring \cite{K41_a,K41_b,K41_c,Frisch}.
Such universal behavior is observed even in systems different from ordinary fluids, such as quantum fluids and supercritical fluids near a critical point \cite{Tsubota_etal_2013,QT_review,Tsatsos_2016,tanogami2021theoretical,skrbek2021phenomenology,tanogami2021van}.
Furthermore, cascade transfer is not limited to such fluid systems, but is also observed even in wave and spin systems \cite{zakharov1992kolmogorov,Nazarenko_2011,fujimoto2012counterflow,fujimoto2012spin,fujimoto2013spin,tsubota2013spin,fujimoto2016direct,rodriguez2021turbulent}. 
We thus conjecture that cascade transfer is a ubiquitous phenomenon.

From this viewpoint, we ultimately aim to establish the concept of a universality class for cascade transfer. 
As a first step toward this end, here we propose a simple model representing one universality class.
We specifically regard the phenomenon as a cooperative phenomenon of unidirectional transport across scales and ask how it emerges from spatially local interactions. 
In contrast, it is nonlocal in most existing cascade models \cite{Frisch,takayasu1995stochastic,mailybaev2015continuous,matsumoto2016one}.

The constructed model is a modified XY model with amplitude fluctuations, in which the spin is regarded as the ``velocity'' of a turbulent field in $d$ dimensions.
We show that the model exhibits an inverse ``energy'' cascade for any $d$, and we calculate the functional form of the velocity correlation function, which corresponds to the non-Kolmogorov energy spectrum $\propto k^{-3}$.
This behavior is quite different from ordinary fluid turbulence even in two dimensions, where the inverse energy cascade inevitably accompanies the enstrophy cascade and the Kolmogorov spectrum.

\paragraph*{Insights into the cascade transfer.}\parhyphen[0pt]
Let us consider the minimum elements required for cascade transfer to occur.
Obviously, nonlinearity is indispensable because the essence of cascade transfer is strong inevitable interference between widely separated length scales.
Furthermore, this nonlinearity must conserve ``energy'' if there is neither injection nor dissipation \cite{Note1}.
To ensure the existence of the ``inertial range,'' the injection and dissipation must act at large (small) and small (large) scales, respectively.
Thus, the minimum elements required for the ``energy'' cascade to occur are (i) nonlinearity that conserves ``energy''; (ii) injection at large (small) scales; and (iii) dissipation at small (large) scales.

\begin{figure}[h]
\includegraphics[width=8.6cm]{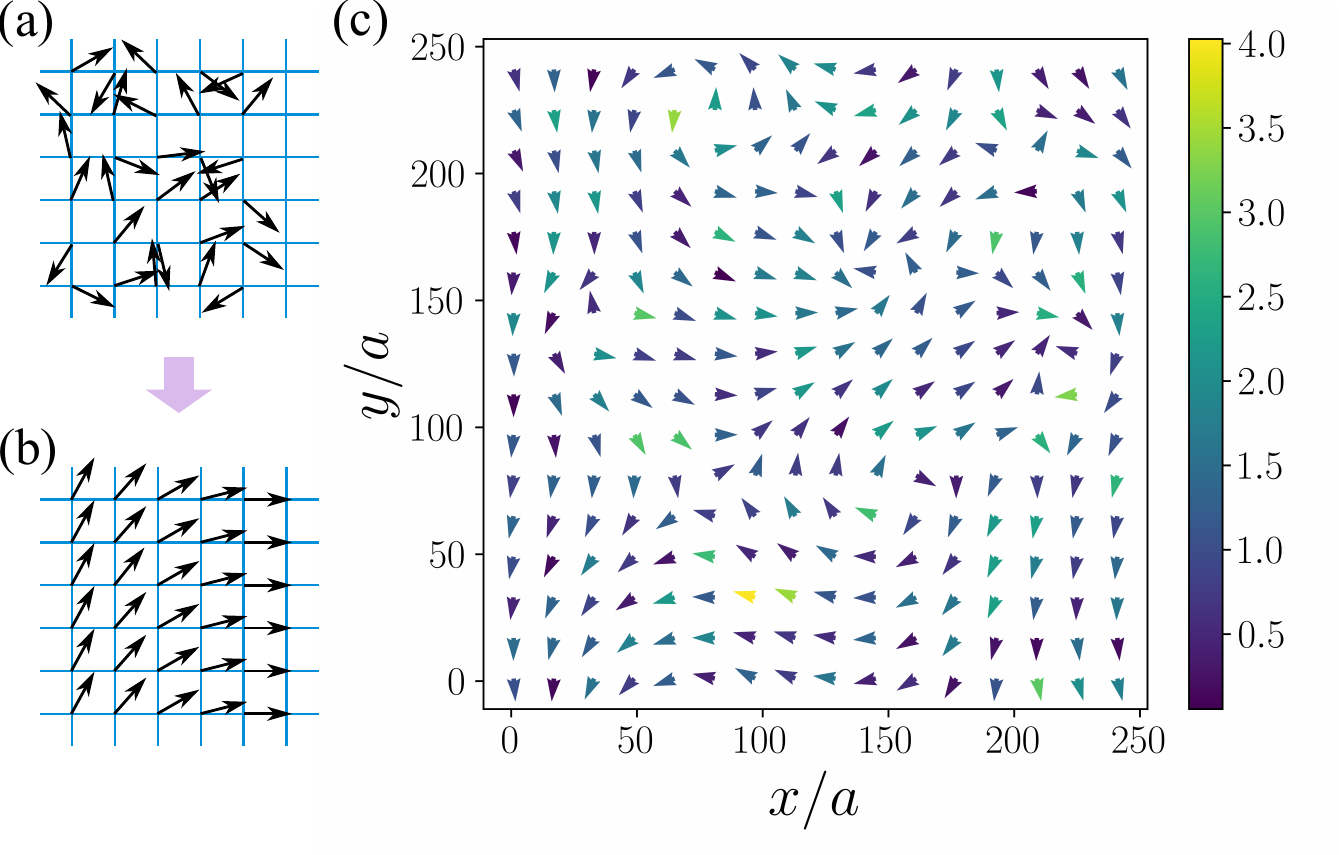}
\caption{(color online). (a) and (b) Schematic illustration of the idea of constructing a simple model. The arrow on each site represents the ``velocity'' of a turbulent field. (c) Snapshot of the steady-state velocity profile of the model with $T=\lambda=1$ and $\gamma=0.001$. The color bar denotes the magnitude of the velocity vector $|{\bm v}_i|$.}
\label{fig:idea}
\end{figure}

We now construct a simple model for cascade transfer by specifying these three elements.
Respecting the ease of the intuitive interpretation of the nonlinear interaction, we consider the two-component ``velocity'' vector ${\bm v}_i$ at each site $i$ on a two-dimensional square lattice.
In the case shown in Figs.\ \ref{fig:idea}(a) and \ref{fig:idea}(b), the ``energy'' $\langle|{\bm v}_i|^2\rangle/2$, where $\langle\cdot\rangle$ denotes the ensemble average, may be localized at small and large scales, respectively.
For the model to evolve from the state shown in Fig.\ \ref{fig:idea}(a) to that shown in Fig.\ \ref{fig:idea}(b) while conserving energy, ``ferromagnetic interactions'' may be suitable nonlinearity.
Because this nonlinear interaction may induce an inverse energy cascade, where the energy is transferred from small to large scales, we must incorporate into the model injection and dissipation terms that act at small and large scales, respectively.
To this end, it may be suitable for the ease of analysis to choose a random force that is white in space and time and a friction dissipation.

\paragraph*{Model.}\parhyphen[0pt]
Let ${\bm v}_i(t):=(v^1_i(t),v^2_i(t))\in\mathbb{R}^2$ be the ``velocity'' at site $i$ of a $d$-dimensional hypercubic lattice.
For simplicity, we consider a hypercubic lattice with $N^d$ vertices and lattice constant $a$ and impose periodic boundary conditions.
The collection of the nearest neighboring sites of $i$ is denoted $B_i$.
The time evolution of $v^a_i$, $a\in\{1,2\}$, is given by the following Langevin equation:
\begin{eqnarray}
\partial_tv^a_i=\lambda\sum_{j\in B_i}\mathsf{R}^{ab}({\bm v}_i)v^b_j-\gamma v^a_i+\sqrt{\epsilon}\xi^a_i,
\label{eq:model}
\end{eqnarray}
where $\mathsf{R}^{ab}({\bm v}_i)$ represents the projection in the direction perpendicular to ${\bm v}_i$:
\begin{eqnarray}
\mathsf{R}^{ab}({\bm v}_i):=\delta^{ab}-\dfrac{v^a_iv^b_i}{|{\bm v}_i|^2}.
\end{eqnarray}
Here, $\lambda>0$ is a coupling constant, $\gamma\ge0$ is a friction coefficient, and $\epsilon>0$ represents the strength of the random force, which is the zero-mean white Gaussian noise that satisfies
\begin{eqnarray}
\langle\xi^a_i(t)\xi^b_j(t')\rangle=\delta^{ab}\delta_{ij}\delta(t-t'),
\end{eqnarray}
and $|{\bm v}_i|^2:=v^c_iv^c_i$.
Here and hereafter, we employ the summation convention for $a, b, c$ that repeated indices in one term are summed over $\{1,2\}$.
A snapshot of the steady-state velocity profile of the model for the case $d=2$ is shown in Fig.\ \ref{fig:idea}(c).
Below, we mainly consider the case of $d=2$, but the extension to any $d$ is straightforward \cite{note2}.

\paragraph*{Basic properties.}\parhyphen[0pt]
Let $|{\bm v}_i|^2/2$ be the ``energy'' at site $i$.
A crucial property of the nonlinear term of the model (\ref{eq:model}) is that the term does not contribute to the energy exchange:
\begin{eqnarray}
v^a_i\left(\lambda\sum_{j\in B_i}\mathsf{R}^{ab}({\bm v}_i)v^b_j\right)=0.
\end{eqnarray}
Therefore, the time evolution of $|{\bm v}_i|^2/2$ is governed only by the dissipation rate $\gamma|{\bm v}_i|^2$ and injection rate $\sqrt{\epsilon}v^c_i\circ\xi^c_i$:
\begin{eqnarray}
\partial_t\dfrac{1}{2}|{\bm v}_i|^2=-\gamma|{\bm v}_i|^2+\sqrt{\epsilon}v^c_i\circ\xi^c_i,
\label{energy conservation}
\end{eqnarray}
where the symbol $\circ$ denotes multiplication in the sense of Stratonovich \cite{gardiner1985handbook}.
Thus, if there is neither injection nor dissipation (i.e., $\epsilon=\gamma=0$), the energy at site $i$, $|{\bm v}_i|^2/2$, is conserved without any averaging.
If $\epsilon>0$ and $\gamma>0$, it follows that $\langle|{\bm v}_i|^2\rangle=2T$ in the steady-state, where we introduced the ``temperature'' as $T:=\epsilon/2\gamma$.

It becomes easier to understand the behavior of the model by introducing the amplitude $A_i$ and the phase $\theta_i$ as ${\bm v}_i=A_i(\cos\theta_i,\sin\theta_i)$.
In terms of $A_i$ and $\theta_i$, (\ref{eq:model}) can be expressed as
\begin{eqnarray}
\partial_tA_i&=&-\gamma A_i+\dfrac{\epsilon}{2A_i}+\sqrt{\epsilon}\xi^A_i,\label{PA_1}\\
A_i\partial_t\theta_i&=&-\lambda\sum_{j\in B_i}A_j\sin(\theta_i-\theta_j)+\sqrt{\epsilon}\xi^\theta_i.\label{PA_2}
\end{eqnarray}
Here, $\xi^A_i:=\xi^1_i\cos\theta_i+\xi^2_i\sin\theta_i$ and $\xi^\theta_i:=-\xi^1_i\sin\theta_i+\xi^2_i\cos\theta_i$, where the multiplication is interpreted in the It\^o sense \cite{gardiner1985handbook}.
Note that (\ref{PA_2}) has the form of the random-bond XY model with asymmetric coupling.
If $A_i$ is frozen uniformly in space, the system exhibits the Kosterlitz-Thouless transition \cite{berezinskii1971destruction,kosterlitz1973ordering,kosterlitz1974critical}.
Therefore, we can say that this model is a modified XY model with amplitude (energy) fluctuations.
We emphasize that, in contrast with the standard XY model, the detailed balance is broken in our model by the amplitude fluctuations.
The absence of the detailed balance is necessary for the cascade transfer to occur in the steady-state.

In the following, we use the property that the energy dissipation and injection act at large and small scales, respectively.
Let $K_i\equiv\ell^{-1}_i$ be the energy injection scale.
Since the injection due to the noise $\xi^a_i$ acts with uniform strength on each Fourier mode, $K_i$ can be defined, for instance, as
\begin{equation}
K_i:=\dfrac{2\pi}{L}\dfrac{1}{N^d}\sum^{N/2}_{n_1=-N/2+1}\cdots\sum^{N/2}_{n_d=-N/2+1}\sqrt{n^2_1+\cdots+n^2_d},
\label{K_i}
\end{equation} 
where $L:=Na$.
The energy injection due to the ``thermal noise'' mainly acts at scales $\ll\ell_i$.
Similarly, let $K_\gamma\equiv\ell^{-1}_\gamma$ be the dissipation scale.
This scale may depend on the friction coefficient $\gamma$ and dissipation rate $\gamma\langle|{\bm v}_i|^2\rangle=\epsilon$.
Therefore, $K_\gamma$ is defined as $K_\gamma:=\gamma^{3/2}\epsilon^{-1/2}$ \cite{bernard1999three,boffetta2010evidence,Boffetta_2012}.
We thus expect that the dissipation is dominant at scales $\gg\ell_\gamma$.
Note that $K_\gamma\rightarrow0$ as $\gamma\rightarrow0$.

\paragraph*{Main result.}\parhyphen[0pt]
Let $\Pi(k)$ be the scale-to-scale energy flux, which represents the energy transfer from scales $> k^{-1}$ to scales $< k^{-1}$.
(The precise definition is given below.)
In the steady-state, $\Pi(k)$ becomes scale independent in the ``inertial range'' $K_\gamma\ll k\ll K_i$:
\begin{eqnarray}
\Pi(k)\simeq-\epsilon<0.
\label{main result_cascade}
\end{eqnarray}
Since $\Pi(k)$ is negative, (\ref{main result_cascade}) states that the model exhibits an inverse energy cascade; i.e., the energy is transferred conservatively and continuously from small to large scales.
Correspondingly, the equal-time correlation function $C({\bm \ell}):=\langle v^c_iv^c_l\rangle$, where ${\bm \ell}:={\bm r}_i-{\bm r}_l$ and ${\bm r}_i$ denotes the position of site $i$, follows a power-law:
\begin{eqnarray}
C({\bm \ell})\sim\dfrac{1}{16}(\lambda a^2)^{-1}\epsilon\ell^2\quad\text{for}\quad \ell_i\ll\ell\ll\ell_\gamma.
\label{main result_correlation}
\end{eqnarray}
From (\ref{main result_correlation}), the one-dimensional energy spectrum $E^{(1D)}(k)$ reads
\begin{equation}
E^{(1D)}(k)\sim C(\lambda a^2)^{-1}\epsilon k^{-3}\quad\text{for}\quad K_\gamma\ll k\ll K_i,
\label{main result_spectrum}
\end{equation}
where $C$ is a positive dimensionless constant.

\paragraph*{Numerical simulation.}\parhyphen[0pt]
We here present the results of numerical simulation for the case $d=2$ \cite{note2}.
Time integration is performed using the simplest discretization method with $\Delta t=0.01$.
The initial value of $v^a_i$ is set as $v^a_i(0)=\sqrt{\epsilon}\Delta W^a_i$, where $\{\Delta W^a_i\}$ denote the independent Wiener processes with variance $\Delta t$. 
The parameter values are chosen as $\lambda=1$, $\epsilon=0.002$, and $\gamma=0.001$, so that $T=1$.
The system size is fixed as $N=1024$ with $a=1$.
In this case, the injection and dissipation scales are estimated as $K_ia\simeq2.41$ and $K_\gamma a\simeq1\times10^{-3}$, respectively.
Note that $K_i$ does not increase but approaches a constant value as $N$ increases.

\begin{figure}[t]
\includegraphics[width=8.6cm]{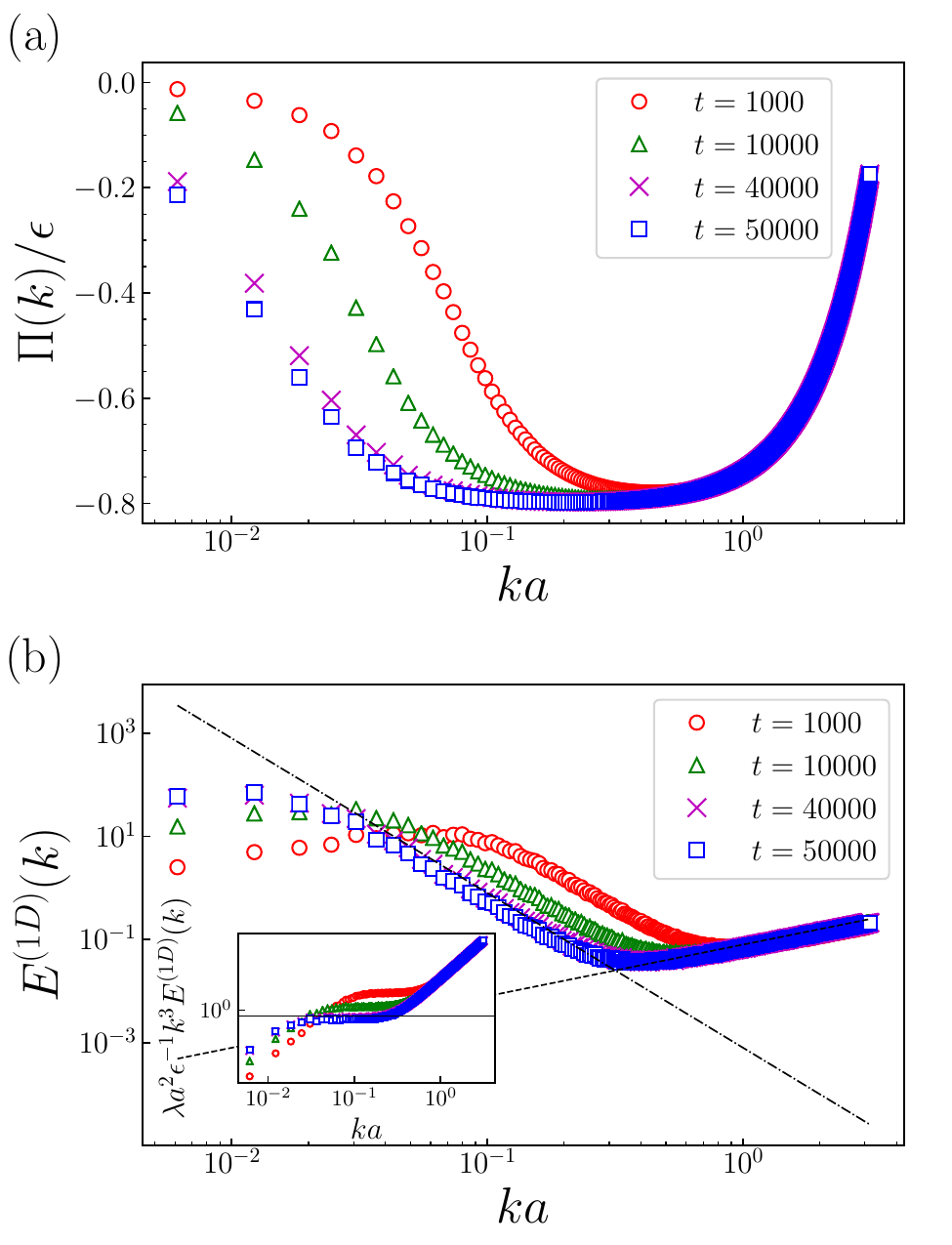}
\caption{(color online). Scale dependence of (a) the scale-to-scale energy flux $\Pi(k)/\epsilon$ and (b) the energy spectrum $E^{(1D)}(k)$ with $T=\lambda=1$ and $\gamma=0.001$ at different times. The dash-dotted and dotted lines represent the power-laws $\propto k^{-3}$ and $\propto k$, respectively. The inset shows the compensated energy spectrum $\lambda a^2\epsilon^{-1}k^3E^{(1D)}(k)$, where the solid line represents $C=1/2$.}
\label{fig:WC_timeevolution_N1024_dt001_Lam1_D0001_gamma0001}
\end{figure}

Figure \ref{fig:WC_timeevolution_N1024_dt001_Lam1_D0001_gamma0001}(a) shows the scale dependence of the scale-to-scale energy flux $\Pi(k)$ at different times.
As expected from the result (\ref{main result_cascade}), $\Pi(k)$ is negative and scale independent in the inertial range $K_\gamma\ll k\ll K_i$.
The magnitude of $\Pi(k)$ in the inertial range is on the order of $\epsilon$, i.e., $\Pi(k)/\epsilon\simeq-1$, which is consistent with (\ref{main result_cascade}).
Furthermore, the scale range over which $\Pi(k)$ is nearly constant extends to larger scales as time increases.
This result also supports that the energy is continuously transferred from small to large scales.
In Fig.\ \ref{fig:WC_timeevolution_N1024_dt001_Lam1_D0001_gamma0001}(b), we plot the one-dimensional energy spectrum $E^{(1D)}(k)$ for the same times as in Fig.\ \ref{fig:WC_timeevolution_N1024_dt001_Lam1_D0001_gamma0001}(a).
In the inertial range, $E^{(1D)}(k)$ follows the power-law $\propto k^{-3}$, which is consistent with the theoretical prediction (\ref{main result_spectrum}).
At scales smaller than the injection scale $K_i$, $E^{(1D)}(k)$ is proportional to $k$.
This result implies that the ``equipartition of energy'' is realized for small scales $\gtrsim K_i$.
We can also confirm the existence of the inverse energy cascade by noting that the spectrum extends to larger scales as time passes.
Note that the range over which $\Pi(k)$ is flat does not exactly correspond to the range over which $E^{(1D)}(k)\propto k^{-3}$.
This discrepancy is similar to that observed in ordinary fluid turbulence \cite{kaneda2003energy}.

\paragraph*{Derivation of the main result.}\parhyphen[0pt]
Let $\hat{v}^a_{\bm k}$ be the discrete Fourier transform of $v^a_i$ with ${\bm k}:=2\pi{\bm n}/L$, where $n^1,n^2\in\{-N/2+1,\cdots,0,1,\cdots,N/2\}$.
We define the low-pass filtering operator by
\begin{eqnarray}
\mathcal{P}^{<K}: {\bm v}_i\mapsto{\bm v}^{<K}_i:=\sum_{|{\bm k}|<K}\hat{{\bm v}}_{{\bm k}}e^{i{\bm k}\cdot{\bm r}_i},
\end{eqnarray}
where $\sum_{|\bm k|<K}$ denotes the sum over all possible ${\bm k}$ that satisfy $|{\bm k}|<K$.
This operator sets to zero all Fourier components with a wavenumber greater than $K$.
By applying this operator to both sides of (\ref{eq:model}) and taking the average, we obtain the low-pass filtered energy balance equation:
\begin{eqnarray}
\partial_t\dfrac{1}{2}\langle|{\bm v}^{<K}_i|^2\rangle&=&-\Pi(K)-\gamma\langle|{\bm v}^{<K}_i|^2\rangle\notag\\
&&+\sqrt{\epsilon}\langle{\bm v}^{<K}_i\circ\bm{\xi}^{<K}_i\rangle,
\end{eqnarray}
where
\begin{eqnarray}
\Pi(K):=-\lambda\left\langle{\bm v}^{<K}_i\cdot\mathcal{P}^{<K}\left[\sum_{j\in B_i}\mathsf{R}(v_i)\cdot{\bm v}_j\right]\right\rangle
\label{Pi_K}
\end{eqnarray}
denotes the scale-to-scale energy flux. 
Note that only $\Pi(K)$ includes the contribution from the Fourier modes with $|{\bm k}|\ge K$ because of the nonlinear interaction.
The dissipation mainly acts at scales $\gg\ell_\gamma$, and it follows that $\gamma\langle|{\bm v}^{<K}_i|^2\rangle\simeq\gamma\langle|{\bm v}^{<K_\gamma}_i|^2\rangle\simeq\gamma\langle|{\bm v}_i|^2\rangle$ for $K_\gamma\ll K$.
Similarly, because the injection mainly acts at scales $\ll\ell_i$, $\langle{\bm v}^{<K}_i\circ\bm{\xi}^{<K}_i\rangle\simeq0$ for $K\ll K_i$.
Therefore, in the steady-state, we obtain
\begin{eqnarray}
\Pi(K)&=&-\gamma\langle|{\bm v}^{<K}_i|^2\rangle+\sqrt{\epsilon}\langle{\bm v}^{<K}_i\circ{\bm \xi}^{<K}_i\rangle\notag\\
&\simeq&-\gamma\langle|{\bm v}_i|^2\rangle\notag\\
&=&-\epsilon<0\quad\text{for}\quad K_\gamma\ll K\ll K_i.
\label{epsilon_def}
\end{eqnarray}
The model thus exhibits the inverse energy cascade; i.e., the energy is transferred conservatively from small to large scales in the ``inertial range'' $K_\gamma\ll K\ll K_i$.
Note that the above argument is essentially the same as that for the two-dimensional fluid turbulence \cite{bernard1999three,boffetta2010evidence,Boffetta_2012,cerbus2017third}.

We now determine the functional form of the energy spectrum.
To this end, we express the energy flux in terms of the velocity correlation function as in the derivation of the Kolmogorov $4/5$-law \cite{Frisch}.
We first note that $\Pi(K)$ can be rewritten as 
\begin{align}
\Pi(K)&=-\left.\partial_t\dfrac{1}{2}\langle|{\bm v}^{<K}_i|^2\rangle\right|_{\mathrm{NL}}\notag\\
&=-\sum_{|{\bm k}|<K}\dfrac{1}{N^2}\sum_{{\bm r}_j-{\bm r}_l}e^{-i{\bm k}\cdot({\bm r}_j-{\bm r}_l)}\left.\partial_t\dfrac{1}{2}\langle v^c_jv^c_l\rangle\right|_{\mathrm{NL}},\notag\\
\label{energy flux rewritten}
\end{align}
where $\left.\partial_t\cdot\right|_{\mathrm{NL}}$ denotes the time evolution due to the nonlinear term.
By taking the continuum limit, (\ref{energy flux rewritten}) can be expressed as
\begin{eqnarray}
\Pi(K)&=&-\int_{|{\bm k}|<K}\dfrac{d^2{\bm k}}{(2\pi)^2}\int d^2{\bm \ell}e^{-i{\bm k}\cdot{\bm \ell}}\epsilon(\ell)\notag\\
&=&-\int^\infty_0Kd\ell J_1(K\ell)\epsilon(\ell).
\label{energy flux continuum limit}
\end{eqnarray}
Here, $J_1$ is the Bessel function of the first kind and we have assumed the homogeneity $\epsilon({\bm \ell}):=\left.\partial_t\langle v^c(\bm{\ell})v^c(\bm{0})\rangle/2\right|_{\mathrm{NL}}=\left.\partial_t\langle v^c(\bm{r}_j)v^c(\bm{r}_l)\rangle/2\right|_{\mathrm{NL}}$ and isotropy $\epsilon({\bm \ell})=\epsilon(\ell)$ with $\bm{\ell}:=\bm{r}_j-\bm{r}_l$.
We now substitute (\ref{energy flux continuum limit}) into the relation (\ref{epsilon_def}) to find
\begin{equation}
\int^\infty_0dxJ_1(x)\epsilon\left(\dfrac{x}{K}\right)\simeq\epsilon\quad\text{for}\quad K_\gamma\ll K\ll K_i.
\label{cascade condition_after approx}
\end{equation}
By taking first the limit $\gamma\rightarrow0$ ($K_\gamma\rightarrow0$) and then the limit $K\rightarrow0$, we obtain, for large $\ell$, \cite{Frisch}
\begin{equation}
\epsilon(\ell)\simeq\epsilon,
\label{epsilon_ell}
\end{equation}
where we have used the identity $\int^\infty_0dxJ_1(x)=1$.
A simple expression for $\epsilon(\ell)$ can be obtained by noting that ${\bm v}_i$ tends to align with $\langle\langle {\bm v}_i\rangle\rangle:=\sum_{j\in B_i}\bm{v}_j/4$ because of the nonlinearity of the model.
In other words, for the angle $\alpha_i$ between $\hat{{\bm v}}_i:={\bm v}_i/|{\bm v}_i|$ and $\langle\langle{\bm v}_i\rangle\rangle/|\langle\langle{\bm v}_i\rangle\rangle|$, we conjecture that $\alpha_i\ll1$ in the steady-state.
Therefore, by assuming that each angle between $\hat{\bm v}_i$ and its nearest neighbor $\hat{\bm v}_j$ is on the order of $\alpha_i\ll1$, we find that
\begin{eqnarray}
\mathsf{R}^{ab}({\bm v}_i)\langle\langle v^b_i\rangle\rangle&=&\langle\langle v^a_i\rangle\rangle-\hat{v}^a_i|\langle\langle{\bm v}_i\rangle\rangle|\cos\alpha_i\notag\\
&\simeq&\langle\langle v^a_i\rangle\rangle-\hat{v}^a_i|\langle\langle{\bm v}_i\rangle\rangle|\notag\\
&\simeq&\langle\langle v^a_i\rangle\rangle-v^a_i+\hat{v}^a_i\left(A_i-\langle\langle A_i\rangle\rangle\right).
\label{approx}
\end{eqnarray}
Since $\{A_i\}$ are independent and identically distributed random variables, we obtain from (\ref{approx}) that
\begin{eqnarray}
&&\partial_t\left.\dfrac{1}{2}\langle v^c_jv^c_l\rangle\right|_{\mathrm{NL}}\notag\\
&=&2\lambda\left[\left\langle v^a_l\mathsf{R}^{ac}({\bm v}_j)\langle\langle v^c_j\rangle\rangle\right\rangle+\left\langle v^a_j\mathsf{R}^{ac}({\bm v}_l)\langle\langle v^c_l\rangle\rangle\right\rangle\right]\notag\\
&\simeq&2\lambda\left[\langle v^c_l\left[\langle\langle v^c_j\rangle\rangle-v^c_j\right]\rangle+\langle v^c_j\left[\langle\langle v^c_l\rangle\rangle-v^c_l\right]\rangle\right],
\label{after refined approx}
\end{eqnarray}
for $|\bm{r}_j-\bm{r}_l|>a$.
Note that $\langle\langle\cdot\rangle\rangle-\cdot$ is the discrete Laplacian.
Therefore, $\epsilon(\ell)$ in (\ref{epsilon_ell}) can be expressed in terms of $C(\ell):=\langle v^c({\bm r}_j)v^c({\bm r}_l)\rangle$:
\begin{equation}
4\lambda a^2\left(\dfrac{\partial^2}{\partial \ell^2}+\dfrac{1}{\ell}\dfrac{\partial}{\partial \ell}\right)C(\ell)\simeq\epsilon.
\end{equation}
It follows from this equation that
\begin{eqnarray}
C(\ell)\sim\dfrac{1}{16}(\lambda a^2)^{-1}\epsilon\ell^2\quad\text{for}\quad \ell_i\ll\ell\ll\ell_\gamma.
\label{correlation function}
\end{eqnarray}
Correspondingly, the asymptotic behavior of the one-dimensional energy spectrum $E^{(1D)}(k)$ in the inertial range reads
\begin{equation}
E^{(1D)}(k)\sim C(\lambda a^2)^{-1}\epsilon k^{-3}\quad\text{for}\quad K_\gamma\ll k\ll K_i,
\label{energy spectrum prediction}
\end{equation}
where $C$ is a dimensionless positive constant.

\paragraph*{Concluding remarks.}\parhyphen[0pt]
One of the fundamental properties of cascades that we have not discussed here is scale locality \cite{Eyink_2005,Eyink_Aluie_2009-1,Aluie_Eyink_2009-2}.
An energy cascade is scale-local if modes that make a significant contribution to energy transfer at each scale are limited to those in the vicinity of that scale.
From the fact that the energy flux and spectrum gradually extend to larger scales as time passes (see Fig.\ \ref{fig:WC_timeevolution_N1024_dt001_Lam1_D0001_gamma0001}), it seems that the inverse cascade is scale-local.
However, a numerical study of scale locality implies that it is not scale-local \cite{note3}, although there remains a problem of how to define the scale locality.
A more detailed study on the scale locality should be carried out in the future.

Interestingly, the behavior of the energy spectrum $E^{(1D)}(k)\propto k^{-3}$ at large scales is also observed in atmospheric turbulence.
In the upper troposphere and lower stratosphere, $E^{(1D)}(k)\propto k^{-5/3}$ at scales between 10 and 500 km while $E^{(1D)}(k)\propto k^{-3}$ at scales between 500 and 3000 km \cite{lilly1983aircraft,nastrom1984kinetic,smithr1994finite,xia2008turbulence,lindborg2009comment,xia2009xia,cerbus2017third}.
We also note that turbulent behavior similar to that of our model is found in so-called spin turbulence \cite{fujimoto2012counterflow,fujimoto2012spin,fujimoto2013spin,tsubota2013spin,fujimoto2016direct,rodriguez2021turbulent} and Fibonacci turbulence \cite{vladimirova2021fibonacci}.
It would thus be interesting to investigate the relationship between these systems and our model.

In conclusion, we constructed a modified XY model in which cascade transfer emerges.
Because this inverse cascade induces the non-Kolmogorov spectrum $E^{(1D)}(k)\propto k^{-3}$, it represents a different universality class from standard fluid turbulence.
We thus hope that our model triggers further investigation of cascade transfer in various systems such as condensed matter, active matter, and other statistical mechanical systems.

TT thanks Takeshi Matsumoto, Shota Shigetomi, Susumu Goto, and Ryo Araki for fruitful discussions.
TT was supported by JSPS KAKENHI Grant No. 20J20079, a Grant-in-Aid for JSPS Fellows.
SS was supported by JSPS KAKENHI (Nos. 17H01148, 19H05795, and 20K20425).

\bibliography{main_text}

\end{document}


\onecolumngrid
\setcounter{equation}{0}
\setcounter{figure}{0}
\setcounter{table}{0}
\makeatletter
\renewcommand{\theequation}{S\arabic{equation}}
\renewcommand{\thefigure}{S\arabic{figure}}
\appendix
\title{Supplemental Material: A Simple XY Model for Cascade Transfer}

\author{Tomohiro Tanogami}
\author{Shin-ichi Sasa}%
\affiliation{%
Department of Physics, Kyoto University, Kyoto 606-8502, Japan
}%





\pacs{Valid PACS appear here}

\maketitle
\onecolumngrid

\subsection{Three-dimensional case}
\begin{figure}[h]
\includegraphics[width=8.6cm]{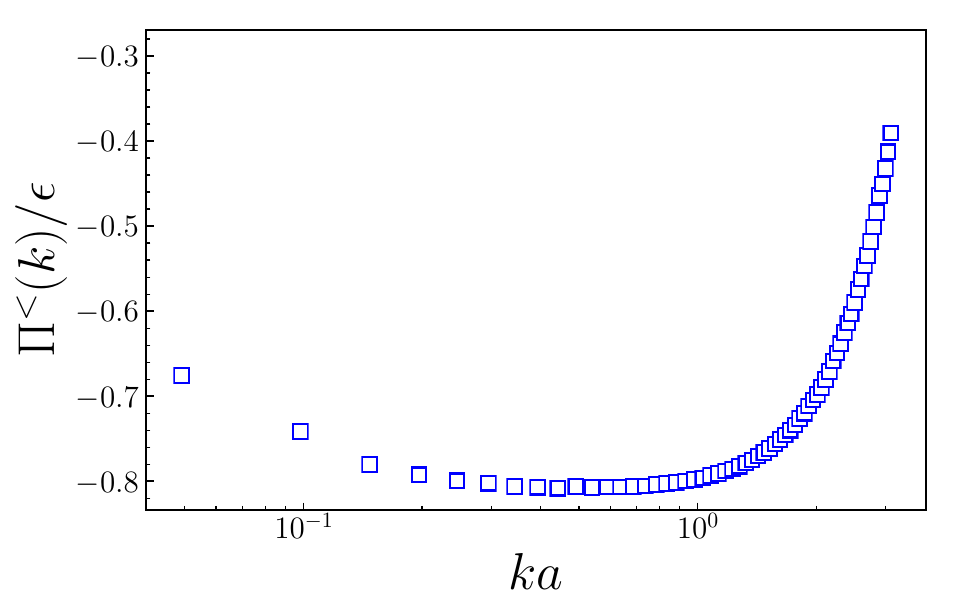}
\includegraphics[width=8.6cm]{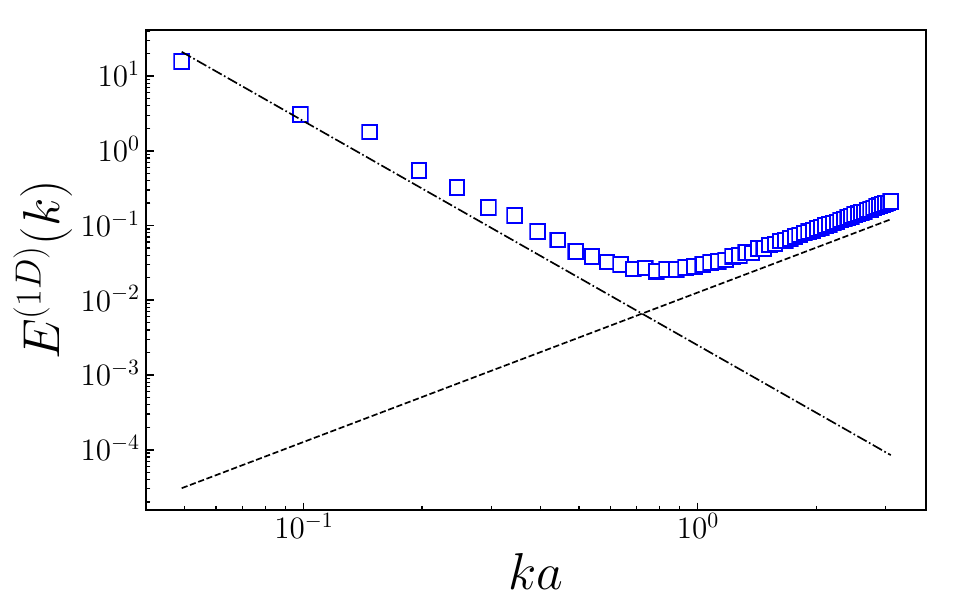}
\caption{Left: the scale-to-scale energy flux $\Pi(k)/\epsilon$. 
Right: the energy spectrum $E^{(1D)}(k)$. The dash-dotted and dotted lines represent the power-laws $\propto k^{-3}$ and $\propto k^2$, respectively.}
\label{fig:3D}
\end{figure}

In this section, we present the numerical result for the three-dimensional case with two-component velocity ${\bf v}_i$.
Even for this case, we can derive the inverse energy cascade with $k^{-3}$ spectrum: for $K_\gamma\ll k\ll K_i$,
\begin{equation}
\Pi(k)\simeq-\epsilon
\label{3D_flux}
\end{equation}
and
\begin{equation}
E^{(1D)}(k)\sim C(\lambda a^2)^{-1}\epsilon k^{-3},
\label{3D_spectrum}
\end{equation}
where $C$ is a positive dimensionless constant.
In Fig.\ \ref{fig:3D}, we plot the scale dependence of the energy flux $\Pi(k)$ and spectrum $E^{(1D)}(k)$.
The parameter values and the system size are chosen as $\lambda=1$, $\epsilon=0.002$, $\gamma=0.001$, and $N=128$ with $a=1$.
In this case, the injection and dissipation scales are estimated as $K_ia\simeq3.02$ and $K_\gamma a\simeq1\times10^{-3}$.
This result is consistent with the prediction (\ref{3D_flux}) and (\ref{3D_spectrum}).
Note that, in this case, $E^{(1D)}(k)\propto k^2$ at small scales, whereas it is proportional to $E^{(1D)}(k)\propto k$ in the two-dimensional case.
This result also implies that the ``equipartition of energy'' is realized for small scales.

\subsection{Scale locality}
In this section, we numerically investigate the scale locality of the inverse energy cascade.
An energy cascade is scale-local if modes that make a significant contribution to energy transfer at each scale are limited to those in the vicinity of that scale.
More precisely, we define the scale locality of the inverse energy cascade for our model as follows \cite{Eyink_2005,Eyink_Aluie_2009-1,Aluie_Eyink_2009-2}.
We first note that the scale-to-scale energy flux $\Pi(K):=-\lambda\langle{\bm v}^{<K}_i\cdot\mathcal{P}^{<K}[\sum_{j\in B_i}\mathsf{R}({\bm v}_i)\cdot{\bm v}_j]\rangle$ has a form of ``velocity'' ${\bm v}^{<K}_i$ times ``force'' $\lambda\mathcal{P}^{<K}[\sum_{j\in B_i}\mathsf{R}({\bm v}_i)\cdot{\bm v}_j]$.
We describe the energy flux as infrared local if $|\lambda\langle{\bm v}^{<Q}_i\cdot\mathcal{P}^{<K}[\sum_{j\in B_i}\mathsf{R}({\bm v}_i)\cdot{\bm v}_j]\rangle|$ and $|\lambda\langle{\bm v}^{<K}_i\cdot\mathcal{P}^{<K}[\sum_{j\in B_i}\mathsf{R}({\bm v}^{<Q}_i)\cdot{\bm v}^{<Q}_j]\rangle|$ gives an asymptotically negligible contribution for $Q\ll K$.
That is, the energy flux satisfies the infrared locality if $|\lambda\langle{\bm v}^{<Q}_i\cdot\mathcal{P}^{<K}[\sum_{j\in B_i}\mathsf{R}({\bm v}_i)\cdot{\bm v}_j]\rangle|$ and $|\lambda\langle{\bm v}^{<K}_i\cdot\mathcal{P}^{<K}[\sum_{j\in B_i}\mathsf{R}({\bm v}^{<Q}_i)\cdot{\bm v}^{<Q}_j]\rangle|$ decay as fast as $(Q/K)^{\alpha}$ with $\alpha>0$ for $Q\ll K$.
Here, we have used the fact that $({\bm v}^{<Q}_i)^{<K}={\bm v}^{<Q}_i$ for $Q\le K$.
Similarly, we describe the energy flux as ultraviolet local if $|\lambda\langle{\bm v}^{<K}_i\cdot\mathcal{P}^{<K}[\sum_{j\in B_i}\mathsf{R}({\bm v}^{>Q}_i)\cdot{\bm v}^{>Q}_j]\rangle|$ decays as fast as $(Q/K)^{-\alpha}$ with $\alpha>0$ for $Q\gg K$.
Here, we do not need to consider $|\lambda\langle({\bm v}^{>Q}_i)^{<K}\cdot\mathcal{P}^{<K}[\sum_{j\in B_i}\mathsf{R}({\bm v}_i)\cdot{\bm v}_j]\rangle|$ because $({\bm v}^{>Q}_i)^{<K}={\bm 0}$ for $Q\ge K$.
\begin{figure}[t]
\includegraphics[width=8.6cm]{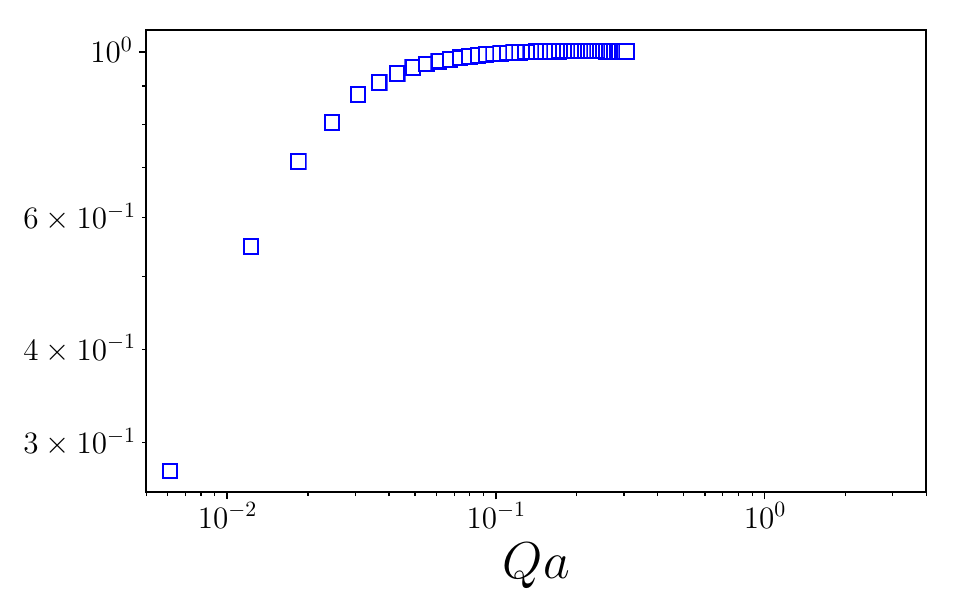}
\includegraphics[width=8.6cm]{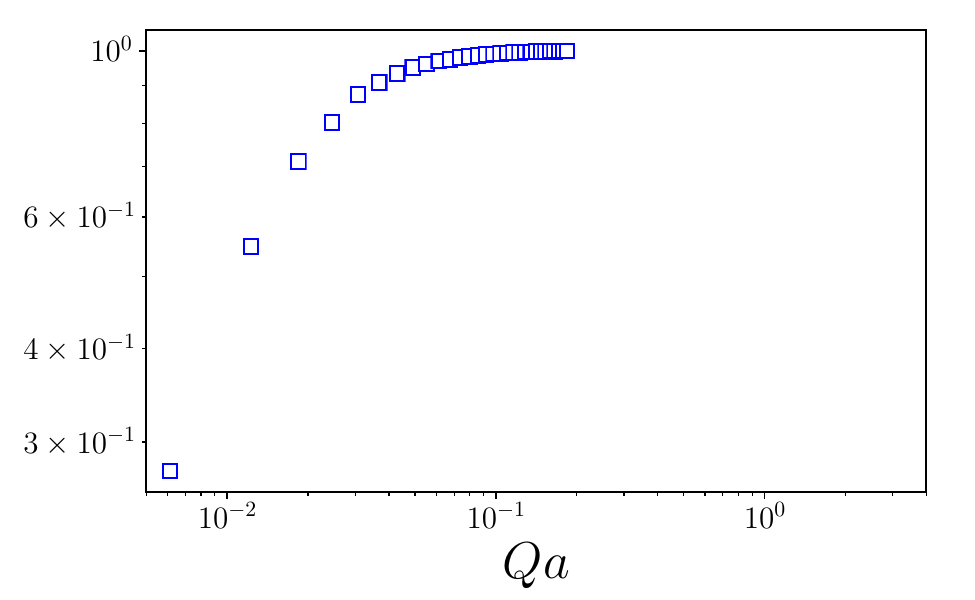}
\caption{$Q$-dependence of $|\lambda\langle{\bm v}^{<Q}_i\cdot\mathcal{P}^{<K}[\sum_{j\in B_i}\mathsf{R}({\bm v}_i)\cdot{\bm v}_j]\rangle/\Pi(K)|$ for $Ka\simeq0.31$ (left) and $Ka\simeq0.18$ (right).
}
\label{fig:IR}
\end{figure}
\begin{figure}[t]
\includegraphics[width=8.6cm]{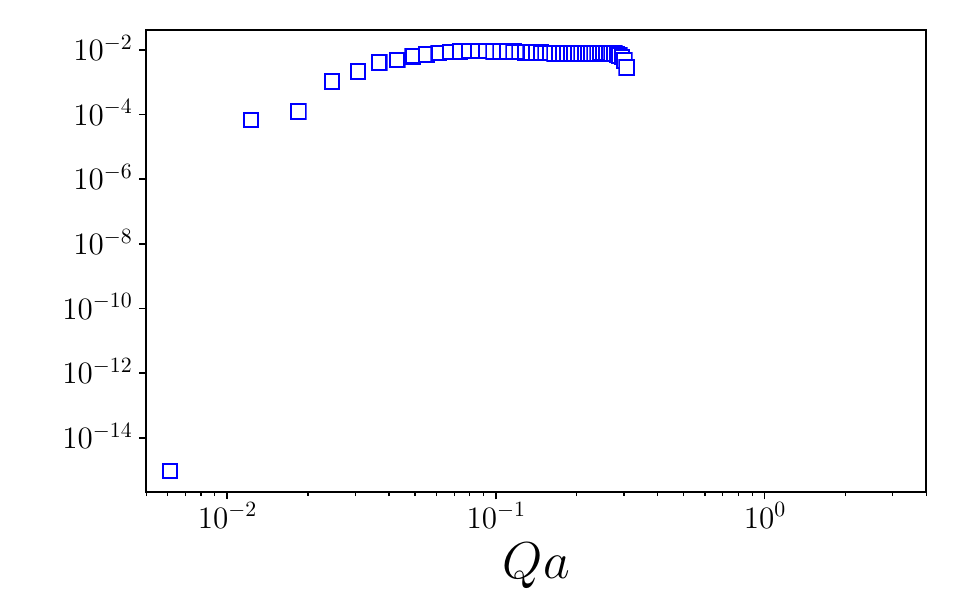}
\includegraphics[width=8.6cm]{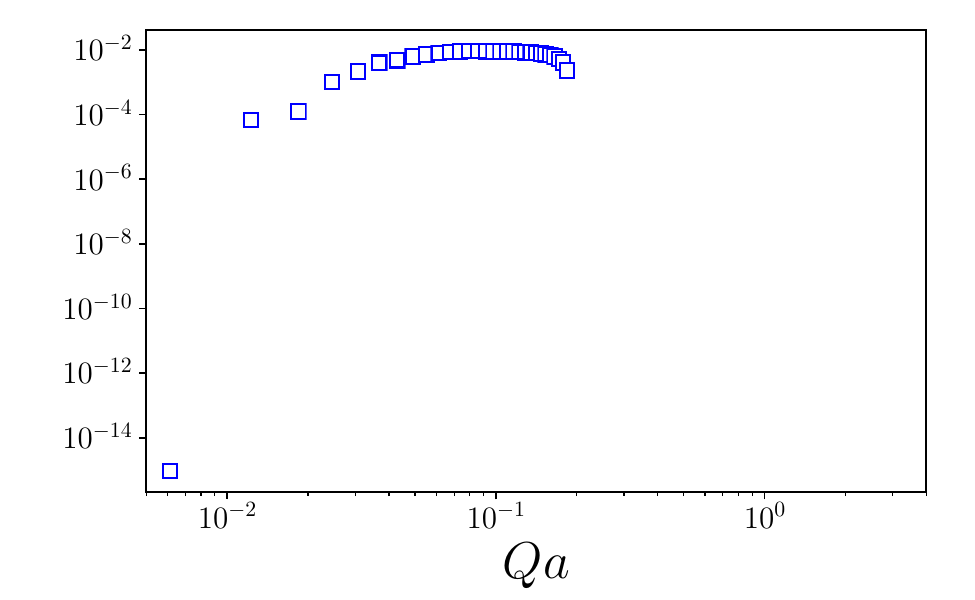}
\caption{$Q$-dependence of $|\lambda\langle{\bm v}^{<K}_i\cdot\mathcal{P}^{<K}[\sum_{j\in B_i}\mathsf{R}({\bm v}^{<Q}_i)\cdot{\bm v}^{<Q}_j]\rangle/\Pi(K)|$ for $Ka\simeq0.31$ (left) and $Ka\simeq0.18$ (right).
}
\label{fig:IR-2}
\end{figure}

We here remark that the energy flux $\Pi(K)$ in our model cannot be scale-local without averaging.
That is, because $\Pi(K)$ is not Galilean invariant, a ${\bm k}={\bm 0}$ mode can directly contribute to the unaveraged energy flux if we boost the flow with a uniform velocity ${\bm U}$, i.e., ${\bm v}_i\mapsto{\bm v}_i+{\bm U}$ for all $i$.
Note that this property is the same as the \textit{unsubtracted flux} for fluid turbulence \cite{Aluie_Eyink_2009-2}.
Even though the unaveraged flux is not scale-local, the averaged flux $\Pi(K)$ may become scale-local because of the cancellation of the large-scale contribution.

We now present the results of numerical simulation.
The parameter values and system size are the same as in the main text: $\lambda=1$, $\epsilon=0.002$, $\gamma=0.001$, and $N=1024$ with $a=1$, so that the injection and dissipation scales are estimated as $K_ia\simeq2.41$ and $K_\gamma a\simeq1\times10^{-3}$.
We first consider the infrared locality.
For the infrared locality, we investigate the $Q$-dependence of the following quantities:
\begin{equation}
\left|\lambda\left\langle{\bm v}^{<Q}_i\cdot\mathcal{P}^{<K}\left[\sum_{j\in B_i}\mathsf{R}({\bm v}_i)\cdot{\bm v}_j\right]\right\rangle\right|,
\label{Tl-lxxx}
\end{equation}
\begin{equation}
\left|\lambda\left\langle{\bm v}^{<K}_i\cdot\mathcal{P}^{<K}\left[\sum_{j\in B_i}\mathsf{R}({\bm v}^{<Q}_i)\cdot{\bm v}^{<Q}_j\right]\right\rangle\right|.
\label{Tl-xlll}
\end{equation}
We calculated these two quantities for $Ka\simeq0.31$ and $Ka\simeq0.18$ ($K=2\pi n/Na$ with $n=50$ and $30$, respectively), which is within the inertial range: $K_\gamma\ll K\ll K_i$.
Figures \ref{fig:IR} and \ref{fig:IR-2} show the $Q$-dependences of these quantities at $t=50000$ normalized by $|\Pi(K)|$ with $Q<K$ for $Ka\simeq0.31$ and $Ka\simeq0.18$, respectively.
Although both quantities decay as $Q\rightarrow0$, they are almost flat in the inertial range $K_\gamma\ll Q<K\ll K_i$.
This result implies that the inverse cascade is not strictly infrared local.
In other words, the contributions to the energy flux from large-scale modes may not be ignored.

\begin{figure}[h]
\includegraphics[width=8.6cm]{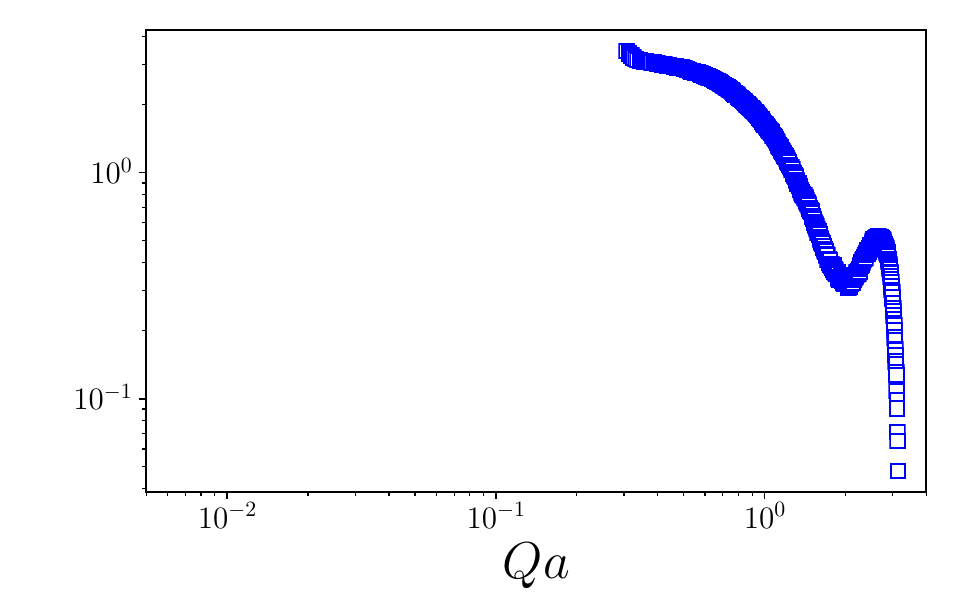}
\includegraphics[width=8.6cm]{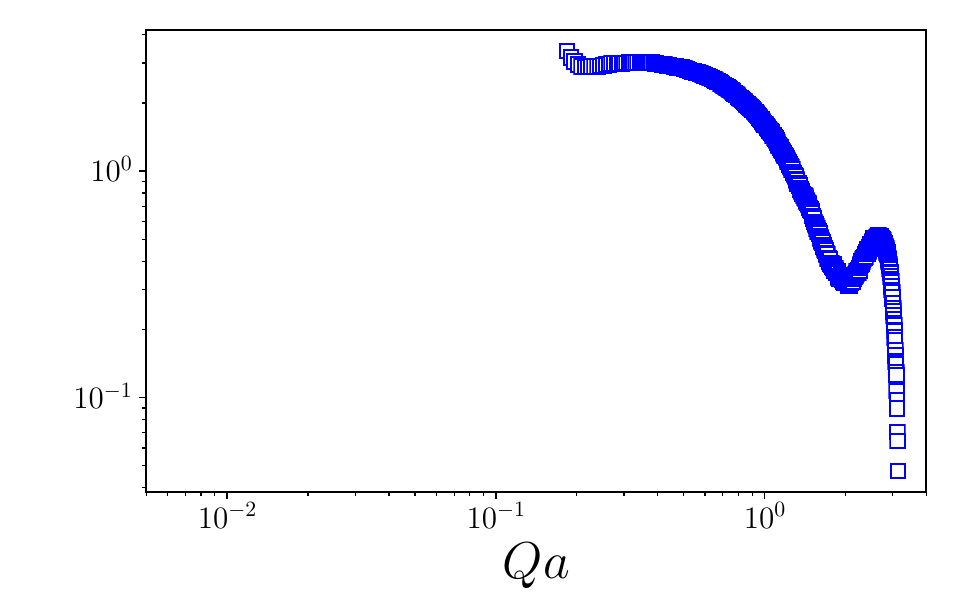}
\caption{$Q$-dependence of $|\lambda\langle{\bm v}^{<K}_i\cdot\mathcal{P}^{<K}[\sum_{j\in B_i}\mathsf{R}({\bm v}^{>Q}_i)\cdot{\bm v}^{>Q}_j]\rangle/\Pi(K)|$ for $Ka\simeq0.31$ (left) and $Ka\simeq0.18$ (right).}
\label{fig:UV}
\end{figure}

For the ultraviolet locality, we investigate the $Q$-dependence of the following quantity:
\begin{equation}
\left|\lambda\left\langle{\bm v}^{<K}_i\cdot\mathcal{P}^{<K}\left[\sum_{j\in B_i}\mathsf{R}({\bm v}^{>Q}_i)\cdot{\bm v}^{>Q}_j\right]\right\rangle\right|.
\label{Tl-xlll}
\end{equation}
The result is shown in Fig.\ \ref{fig:UV}.
As in the case of the infrared locality, $|\lambda\langle{\bm v}^{<K}_i\cdot\mathcal{P}^{<K}[\sum_{j\in B_i}\mathsf{R}({\bm v}^{>Q}_i)\cdot{\bm v}^{>Q}_j]\rangle/\Pi(K)|$ does not decay rapidly in the inertial range $K_\gamma\ll K<Q\ll K_i$.
Therefore, small-scale modes may contribute significantly to the energy flux.


\bibliography{main_text}